\documentclass{article}

\usepackage{color}
\usepackage{amsmath, amssymb, amsthm}

\newcommand{\RM}{\mathbb{R}}
\newcommand{\ZM}{\mathbb{Z}}

\newcommand{\CM}{\mathbb{C}}
\newtheorem{theorem}{Theorem}
\newtheorem{lemma}{Lemma} 
\newtheorem{prop}{Proposition} 
\newtheorem{cor}{Corollary}

\newtheorem*{proof*}{Proof}

\newcommand{\xvec}{\ensuremath{\boldsymbol{x}}}

\newcommand{\evec}{\ensuremath{\boldsymbol{e}}}

\newcommand{\kvec}{\ensuremath{\boldsymbol{k}}}
\newcommand{\wvec}{\ensuremath{\boldsymbol{w}}}

\begin{document}

\title{{\bf Mahler/Zeta Correspondence}
\vspace{15mm}}

\author{Takashi KOMATSU \\
Math. Research Institute Calc for Industry \\
Minami, Hiroshima, 732-0816, Japan \\ 
e-mail: ta.komatsu@sunmath-calc.co.jp 
\\ \\
Norio KONNO \\
Department of Applied Mathematics, Faculty of Engineering \\ 
Yokohama National University \\
Hodogaya, Yokohama, 240-8501, Japan \\
e-mail: konno-norio-bt@ynu.ac.jp 
\\ \\
Iwao SATO \\ 
Oyama National College of Technology \\
Oyama, Tochigi, 323-0806, Japan \\ 
e-mail: isato@oyama-ct.ac.jp 
\\ \\
Shunya TAMURA \\
Graduate School of Science and Engineering \\
Yokohama National University \\ 
Hodogaya, Yokohama, 240-8501, Japan \\ 
e-mail: tamura-shunya-kj@ynu.jp 
}

\date{\empty }

\maketitle

\vspace{50mm}


\vspace{20mm}


%









\clearpage

\begin{abstract}
The Mahler measure was introduced by Mahler in the study of number theory. It is known that the Mahler measure appears in different areas of mathematics and physics. On the other hand, we have been investigated a new class of zeta functions for various kinds of walks including quantum walks by a series of our previous work on ``Zeta Correspondence". The quantum walk is a quantum counterpart of the random walk. In this paper, we present a new relation between the Mahler measure and our zeta function for quantum walks. Firstly we consider this relation in the case of one-dimensional quantum walks. Afterwards we deal with higher-dimensional quantum walks. For comparison with the case of the quantum walk, we also treat the case of higher-dimensional random walks. Our results bridge between the Mahler measure and the zeta function via quantum walks for the first time.
\end{abstract}

\vspace{10mm}

\begin{small}
\end{small}
\begin{small}
\par\noindent
{\bf Abbr. title:} Mahler/Zeta Correspondence
\end{small}

\vspace{10mm}

\section{Introduction \label{sec01}}
We have been investigated a new class of zeta functions for many kinds of walks including the {\em quantum walk} (QW) and the {\em random walk} (RW) by a series of our previous work on ``Zeta Correspondence" in \cite{K1, K2, K3, K4, K5, K7, K6}. The QW can be interpreted as a quantum counterpart of RW. As for QW, see \cite{Konno2008, ManouchehriWang, Portugal, Venegas} and as for RW, see \cite{Konno2009, Norris, Spitzer}, for examples. Guttmann and Rogers \cite{GR} studied the connection between spanning tree generating functions and lattice Green functions on various lattices such as square, cubic, triangle, honeycomb lattices, by using known results for the logarithmic Mahler measure $m(f)$ of a Laurant polynomial $f$. Sometimes we simply refer to $m(f)$ as the {\em Mahler measure}. This measure was introduced by Mahler \cite{M} in the study of transcendental numbers. It is known that the Mahler measure appears in different areas of mathematics and physics, such as number theory, dynamical systems, Calabi-Yau differential equations, dimer models (see \cite{G,GR}, for example). Following their argument in \cite{GR}, we present a relation between the Mahler measure and our zeta function via QWs, so we call this relationship ``Mahler/Zeta Correspondence" for short.  Specifically, we first deal with this relation in the case of one-dimensional QWs including the Hadamard walk. After that, we treat the higher-dimensional Grover walk. Both the Hadamard and Grover walks are well-investigated models in the study of the QW. For comparison with the case of the QW, we also give some results on the relation between Mahler measures and zeta functions for $d$-dimensional RWs. Our results bridge between the Mahler measure and the zeta function research fields via QWs for the first time. To clarify the mathematical structure of the QW by the Mahler measure will be useful for application to quantum information theory.

The rest of this paper is organized as follows. In Section \ref{sec02}, we briefly review ``Walk/Zeta Correspondence" investigated in \cite{K2}. Section \ref{sec07m} presents the definition of the Mahler measure. In Section \ref{sec03new}, we deal with a relation between the zeta function (introduced in Section \ref{sec02}) and the Mahler measure (introduced in Section \ref{sec07m}) for the two-state QW in one dimension. Section \ref{sec07} is devoted to the case of the higher-dimensional QW, more precisely, the Grover walk. In Section \ref{secmulti}, we consider higher-dimensional RWs in order to compare QW with RW. Section \ref{sec09} concludes our results.

\section{Walk/Zeta Correspondence \label{sec02}}
For the convenience of readers, we give a brief overview of ``Walk/Zeta Correspondence" in our previous work \cite{K2}. This correspondence is one of the fundamental tools in this paper.

First we introduce the following notation: $\mathbb{Z}$ is the set of integers, $\mathbb{Z}_{\ge}$ is the set of non-negative integers, $\mathbb{Z}_{>}$ is the set of positive integers, $\mathbb{R}$ is the set of real numbers, and $\mathbb{C}$ is the set of complex numbers. Moreover, $T^d_N$ denotes the {\em $d$-dimensional torus} with $N^d$ vertices, where $d, \ N \in \mathbb{Z}_{>}$. Remark that $T^d_N = (\mathbb{Z} \ \mbox{mod}\ N)^{d}$.

Following \cite{K2} in which Walk/Zeta Correspondence on $T^d_N$ was investigated, we treat our setting for $2d$-state discrete time walk with a nearest-neighbor jump on $T^d_N$.

The discrete time walk is defined by using a {\em shift operator} and a {\em coin matrix} which will be mentioned below. Let $f : T^d_N \longrightarrow \mathbb{C}^{2d}$. For $j = 1,2,\ldots,d$ and $\xvec \in T^d_N$, the shift operator $\tau_j$ is defined by $(\tau_j f)(\xvec) = f(\xvec-\evec_{j})$, 
where $\{ \evec_1,\evec_2,\ldots,\evec_d \}$ denotes the standard basis of $T^d_N$. The {\em coin matrix} $A=[a_{ij}]_{i,j=1,2,\ldots,2d}$ is a $2d \times 2d$ matrix with $a_{ij} \in \mathbb{C}$ for $i,j =1,2,\ldots,2d$. If $a_{ij} \in [0,1]$ and $\sum_{i=1}^{2d} a_{ij} = 1$ for any $j=1,2, \ldots, 2d$, then the walk is a {\em correlated random walk} (CRW). We should remark that, in particular, when $a_{i1} = a_{i2} = \cdots = a_{i 2d}$ for any $i=1,2, \ldots, 2d$, this CRW becomes a RW. If $A$ is unitary, then the walk is a QW. So our class of walks contains RW, CRW, and QW as special models. 

To describe the evolution of the walk, we decompose the $2d \times 2d$ coin matrix $A$ as
\begin{align*}
A=\sum_{j=1}^{2d} P_{j} A,
\end{align*}
where $P_j$ denotes the orthogonal projection onto the one-dimensional subspace $\mathbb{C}\eta_j$ in $\mathbb{C}^{2d}$. Here $\{\eta_1,\eta_2, \ldots, \eta_{2d}\}$ denotes a standard basis on $\mathbb{C}^{2d}$.

The discrete time walk associated with the coin matrix $A$ on $T^d_N$ is determined by the $2d N^d \times 2d N^d$ matrix
\begin{align}
M_A=\sum_{j=1}^d \Big( P_{2j-1} A \tau_{j}^{-1} + P_{2j} A \tau_{j} \Big).
\label{unitaryop1}
\end{align}
The state at time $n \in \mathbb{Z}_{\ge}$ and location $\xvec \in T^d_N$ can be expressed by a $2d$-dimensional vector:
\begin{align*}
\Psi_{n}(\xvec)= {}^T
\left[
\Psi^{1}_{n}(\xvec), \Psi^{2}_{n}(\xvec), \ldots , \Psi^{2d}_{n}(\xvec) 
\right]
\in \mathbb{C}^{2d},
\end{align*}
where $T$ is the transposed operator. For $\Psi_n : T^d_N \longrightarrow \mathbb{C}^{2d} \ (n \in \mathbb{Z}_{\geq})$, Eq. \eqref{unitaryop1} gives the evolution of the walk as follows.
\begin{align}
\Psi_{n+1}(\xvec) \equiv (M_A \Psi_{n})(\xvec)=\sum_{j=1}^{d}\Big(P_{2j-1}A\Psi_{n}(\xvec+\evec_j)+P_{2j}A\Psi_{n}(\xvec-\evec_j)\Big).
\label{reunitaryop1}
\end{align} 
This equation means that the walker moves at each step one unit to the $- x_j$-axis direction with matrix $P_{2j-1}A$ or one unit to the $x_j$-axis direction with matrix $P_{2j}A$ for $j=1,2, \ldots, d$. Moreover, for $n \in \ZM_{>}$ and $\xvec = (x_1, x_2, \ldots, x_d) \in T^d_N$, the $2d \times 2d$ matrix $\Phi_n (x_1, x_2, \ldots, x_d)$ is given by 
\begin{align*}
\Phi_n (x_1, x_2, \ldots, x_d) = \sum_{\ast} \Xi_n \left(l_1,l_2, \ldots , l_{2d-1}, l_{2d} \right),
\end{align*} 
where the $2d \times 2d$ matrix $\Xi_n \left(l_1,l_2, \ldots , l_{2d-1}, l_{2d} \right)$ is the sum of all possible paths in the trajectory of $l_{2j-1}$ steps $- x_j$-axis direction and  $l_{2j}$ steps $x_j$-axis direction and $\sum_{\ast}$ is the summation over $\left(l_1,l_2, \ldots , l_{2d-1}, l_{2d} \right) \in (\ZM_{\ge})^{2d}$ satisfying 
\begin{align*}
l_1 + l_2 + \cdots + l_{2d-1} + l_{2d} = n, \qquad x_j = - l_{2j-1} + l_{2j} \quad (j=1,2, \ldots, d).
\end{align*} 
Here we put
\begin{align*}
\Phi_0 (x_1, x_2, \ldots, x_d)
= \left\{ 
\begin{array}{ll}
I_{2d} & \mbox{if $(x_1, x_2, \ldots, x_d) = (0, 0, \ldots, 0)$, } \\
O_{2d} & \mbox{if $(x_1, x_2, \ldots, x_d) \not= (0, 0, \ldots, 0)$},
\end{array}
\right.
\end{align*}
where $I_n$ is the $n \times n$ identity matrix and $O_n$ is the $n \times n$ zero matrix. Then, for the walk starting from $(0,0, \ldots, 0)$, we obtain  
\begin{align*}
\Psi_n (x_1, x_2, \ldots, x_d) = \Phi_n (x_1, x_2, \ldots, x_d) \Psi_0 (0, 0, \ldots, 0) \qquad (n \in \mathbb{Z}_{\ge}).
\end{align*} 
We call $\Phi_n (\xvec) = \Phi_n (x_1, x_2, \ldots, x_d)$ {\em matrix weight} at time $n \in \mathbb{Z}_{\ge}$ and location $\xvec \in T^d_N$ starting from ${\bf 0} = (0,0, \ldots, 0)$. When we consider the walk on not $T^d_N$ but $\ZM^d$, we add the superscript ``$(\infty)$" to the notation like $\Psi^{(\infty)}$ and $\Xi^{(\infty)}$.

This type is {\em moving} shift model called {\em M-type} here. Another type is {\em flip-flop} shift model called {\em F-type} whose coin matrix is given by 
\begin{align*}
A^{(f)} = \left( I_{d} \otimes \sigma \right) A,
\end{align*} 
where $\otimes$ is the tensor product and
\begin{align*}
\sigma 
=
\begin{bmatrix}
0 & 1 \\ 
1 & 0 
\end{bmatrix} 
. 
\end{align*}
The F-type model is also important, since it has a central role in the Konno-Sato theorem \cite{KS}, for example. When we distinguish $A$ (M-type) from $A^{(f)}$ (F-type), we write $A$ by $A^{(m)}$.

The measure $\mu_n (\xvec)$ at time $n \in \mathbb{Z}_{\ge}$ and location $\xvec \in T^d_N$ is defined by
\begin{align*}
\mu_n (\xvec) = \| \Psi_n (\xvec) \|_{\mathbb{C}^{2d}}^p = \sum_{j=1}^{2d}|\Psi_n^{j}(\xvec)|^p,
\end{align*} 
where $\|\cdot\|_{\mathbb{C}^{2d}}^p$ denotes the standard $p$-norm on $\mathbb{C}^{2d}$. As for RW and QW, we take $p=1$ and $p=2$, respectively. Then RW and QW satisfy 
\begin{align*}
\sum_{\xvec \in T_N^d} \mu_n (\xvec) = \sum_{\xvec \in T_N^d} \mu_0 (\xvec), 
\end{align*} 
for any time $n \in \ZM_{>}$. 

To consider the zeta function, we use the Fourier analysis. To do so, we introduce the following notation: $\mathbb{K}_N = \{ 0,1, \ldots, N-1 \}$ and $\widetilde{\mathbb{K}}_N = \{ 0 ,2 \pi/N, \ldots, 2 \pi (N-1)/N \}$.

For $f : \mathbb{K}_N^d \longrightarrow \mathbb{C}^{2d}$, the Fourier transform of the function $f$, denoted by $\widehat{f}$, is defined by the sum
\begin{align}
\widehat{f}(\kvec) = \frac{1}{N^{d/2}} \sum_{\xvec \in \mathbb{K}_N^d} e^{- 2 \pi i \langle \xvec, \kvec \rangle /N} \ f(\xvec),
\label{yoiko01}
\end{align}
where $\kvec=(k_1,k_2,\ldots,k_{d}) \in \mathbb{K}_N^d$. Here $\langle \xvec,  \kvec \rangle$ is the canonical inner product of $\mathbb{R}^d$, i.e., $\langle \xvec,  \kvec \rangle = \sum_{j=1}^{d} x_j k_j$. Then we see that $\widehat{f} : \mathbb{K}_N^d \longrightarrow \mathbb{C}^{2d}$. Moreover, we should remark that 
\begin{align}
f(\xvec) = \frac{1}{N^{d/2}} \sum_{\kvec \in \mathbb{K}_N^d} e^{2 \pi i \langle \xvec, \kvec \rangle /N} \ \widehat{f}(\kvec),
\label{yoiko02}
\end{align}
where $\xvec =(x_1,x_2,\ldots,x_{d}) \in \mathbb{K}_N^d$. By using 
\begin{align}
\widetilde{k}_j = \frac{2 \pi k_j}{N} \in \widetilde{\mathbb{K}}_N, \quad \widetilde{\kvec}=(\widetilde{k}_1,\widetilde{k}_2,\ldots,\widetilde{k}_{d}) \in \widetilde{\mathbb{K}}_N^d, 
\label{kantan01}
\end{align}
we can rewrite Eqs. \eqref{yoiko01} and \eqref{yoiko02} in the following way: 
\begin{align*}
\widehat{g}(\widetilde{\kvec}) 
= \frac{1}{N^{d/2}} \sum_{\xvec \in \mathbb{K}_N^d} e^{- i \langle \xvec, \widetilde{\kvec} \rangle} \ g(\xvec),
\qquad 
g(\xvec) 
= \frac{1}{N^{d/2}} \sum_{\widetilde{\kvec} \in \widetilde{\mathbb{K}}_N^d} e^{i \langle \xvec, \widetilde{\kvec} \rangle} \ \widehat{g}(\widetilde{\kvec}),
\end{align*}
for $g : \mathbb{K}_N^d \longrightarrow \mathbb{C}^{2d}$ and $\widehat{g} : \widetilde{\mathbb{K}}_N^d \longrightarrow \mathbb{C}^{2d}$. In order to take a limit $N \to \infty$, we introduced the notation given in Eq. \eqref{kantan01}. We should note that as for the summation, we sometimes write ``$\kvec \in \mathbb{K}_N^d$" instead of ``$\widetilde{\kvec} \in \widetilde{\mathbb{K}}_N^d$". From the Fourier transform and Eq. \eqref{reunitaryop1}, we have
\begin{align*}
\widehat{\Psi}_{n+1}(\kvec)=\widehat{M}_A(\kvec)\widehat{\Psi}_n(\kvec),
\end{align*}
where $\Psi_n : T_N^d \longrightarrow \mathbb{C}^{2d}$ and $2d \times 2d$ matrix $\widehat{M}_A(\kvec)$ is determined by
\begin{align*}                          
\widehat{M}_A(\kvec)=\sum_{j=1}^{d} \Big( e^{2 \pi i k_j/N} P_{2j-1} A + e^{-2 \pi i k_j /N} P_{2j} A \Big). 
\end{align*}
By using notations in Eq. \eqref{kantan01}, we get 
\begin{align}                          
\widehat{M}_A(\widetilde{\kvec})=\sum_{j=1}^{d} \Big( e^{i \widetilde{k}_j} P_{2j-1} A + e^{-i \widetilde{k}_j} P_{2j} A \Big). 
\label{migiyoshi01}
\end{align}
Next we will consider the following eigenvalue problem for $2d N^d \times 2d N^d$ matrix $M_A$: 
\begin{align}
\lambda \Psi = M_A \Psi, 
\label{yoiko03}
\end{align}
where $\lambda \in \mathbb{C}$ is an eigenvalue and $\Psi (\in \mathbb{C}^{2d N^d})$ is the corresponding eigenvector. Noting that Eq. \eqref{yoiko03} is closely related to Eq. \eqref{reunitaryop1}, we see that Eq. \eqref{yoiko03} is rewritten as 
\begin{align}
\lambda \Psi (\xvec) = (M_A \Psi)(\xvec) = \sum_{j=1}^{d}\Big(P_{2j-1} A \Psi (\xvec+\evec_j)+P_{2j} A \Psi (\xvec-\evec_j)\Big),
\label{yoiko04}
\end{align} 
for any $\xvec \in \mathbb{K}_N^d$. From the Fourier transform and Eq. \eqref{yoiko04}, we obtain
\begin{align*}
\lambda \widehat{\Psi} (\kvec) = \widehat{M}_A (\kvec) \widehat{\Psi} (\kvec), 
\end{align*}
for any $\kvec \in \mathbb{K}_N^d$. Then the characteristic polynomials of $2d \times 2d$ matrix $\widehat{M}_A(\kvec)$ for fixed $\kvec (\in \mathbb{K}_N^d)$ is 
\begin{align}                          
\det \Big( \lambda I_{2d} - \widehat{M}_A (\kvec) \Big) = \prod_{j=1}^{2d} \Big( \lambda - \lambda_{j} (\kvec) \Big),
\label{yoiko06}
\end{align}
where $\lambda_{j} (\kvec)$ are eigenvalues of $\widehat{M}_A (\kvec)$. Similarly, the characteristic polynomials of $2d N^d \times 2d N^d$ matrix $\widehat{M}_A$ is 
\begin{align*}                          
\det \Big( \lambda I_{2d N^d} - \widehat{M}_A  \Big) = \prod_{j=1}^{2d} \prod_{\kvec \in \mathbb{K}_N^d} \Big( \lambda - \lambda_{j} (\kvec) \Big).
\end{align*}
Therefore, by taking $\lambda = 1/u$, we get the following key result.
\begin{align}                          
\det \Big( I_{2d N^d} - u M_A  \Big) 
= \det \Big( I_{2d N^d} - u \widehat{M}_A  \Big) = \prod_{j=1}^{2d} \prod_{\kvec \in \mathbb{K}_N^d} \Big( 1 - u \lambda_{j} (\kvec) \Big).
\label{keyresult01lemma}
\end{align}
We should note that for fixed $\kvec (\in \mathbb{K}_N^d)$, eigenvalues of  $2d \times 2d$ matrix $\widehat{M}_A (\kvec)$ are expressed as 
\begin{align*}
{\rm Spec} ( \widehat{M}_A (\kvec)) = \left\{  \lambda_{j} (\kvec) \ | \ j = 1, 2, \ldots, 2d \right\}. 
\end{align*}
Moreover, eigenvalues of $2d N^d \times 2d N^d$ matrix not only $\widehat{M}_A$ but also $M_A$ are expressed as 
\begin{align*}
{\rm Spec} ( \widehat{M}_A) = {\rm Spec} (M_A)= \left\{  \lambda_{j} (\kvec) \ | \ j = 1, 2, \ldots, 2d, \ \kvec \in \mathbb{K}_N^d \right\}. 
\end{align*}

By using notations in Eq. \eqref{kantan01} and Eq. \eqref{yoiko06}, we see that for fixed $\kvec (\in \mathbb{K}_N^d)$,  
\begin{align}                          
\det \Big( I_{2d} - u \widehat{M}_A (\widetilde{\kvec}) \Big) = \prod_{j=1}^{2d} \Big( 1 - u \lambda_{j} (\widetilde{\kvec}) \Big).
\label{yoiko08}
\end{align}
Furthermore, Eq. \eqref{migiyoshi01} gives the following important formula.
\begin{align*}                          
\det \Big( I_{2d} - u \widehat{M}_A (\widetilde{\kvec}) \Big) = \det \left(I_{2d} - u \times \sum_{j=1}^{d} \Big( e^{i \widetilde{k}_j} P_{2j-1} A + e^{-i \widetilde{k}_j} P_{2j} A \Big) \right). 
\end{align*}

In this setting, we define the {\em walk-type zeta function} by 
\begin{align}
\overline{\zeta} \left(A, T^d_N, u \right) = \det \Big( I_{2d N^d} - u M_A \Big)^{-1/N^d}.
\label{satosan01}
\end{align}
We should remark that our walk is defined on the ``site" $\xvec (\in T^d_N)$. On the other hand, the walk in \cite{K1} is defined on the ``arc" (i.e., oriented edge). However, both of the walks are the same for the torus case.

By Eqs. \eqref{keyresult01lemma}, \eqref{yoiko08} and \eqref{satosan01}, we get 
\begin{align*}
\overline{\zeta} \left(A, T^d_N, u \right) ^{-1}
=
\exp \left[ \frac{1}{N^d} \sum_{\widetilde{\kvec} \in \widetilde{\mathbb{K}}_N^d} \log \left\{ \det \Big( I_{2d} - u \widehat{M}_A (\widetilde{\kvec}) \Big) \right\} \right].
\end{align*}
Sometimes we write $\sum_{\kvec \in \mathbb{K}_N^d}$ instead of $\sum_{\widetilde{\kvec} \in \widetilde{\mathbb{K}}_N^d}$. Noting $\widetilde{k_j} = 2 \pi k_j/N \ (j=1,2, \ldots, d)$ and taking a limit as $N \to \infty$, we show
\begin{align*}
\lim_{N \to \infty} \overline{\zeta} \left(A, T^d_N, u \right) ^{-1}
=
\exp \left[ \int_{[0,2 \pi)^d} \log \left\{ \det \Big( I_{2d} - u \widehat{M}_A \left( \Theta^{(d)} \right) \Big) \right\} d \Theta^{(d)}_{unif} \right],
\end{align*}
if the limit exists for a suitable range of $u \in \RM$. We should note that when we take a limit as $N \to \infty$, we assume that the limit exists throughout this paper. Here $\Theta^{(d)} = (\theta_1, \theta_2, \ldots, \theta_d) (\in [0, 2 \pi)^d)$ and $d \Theta^{(d)}_{unif}$ denotes the uniform measure on $[0, 2 \pi)^d$, that is,
\begin{align*}
d \Theta^{(d)}_{unif} = \frac{d \theta_1}{2 \pi } \cdots \frac{d \theta_d}{2 \pi }.
\end{align*}
Then the following result was obtained.
\begin{theorem}[Komatsu, Konno and Sato \cite{K2}]
\begin{align*}
\overline{\zeta} \left(A, T^d_N, u \right) ^{-1}
&= \exp \left[ \frac{1}{N^d} \sum_{\widetilde{\kvec} \in \widetilde{\mathbb{K}}_N^d} \log \left\{ \det \Big( F(\widetilde{\kvec}, u) \Big) \right\} \right],
\\
\lim_{N \to \infty} \overline{\zeta} \left(A, T^d_N, u \right) ^{-1}
&=
\exp \left[ \int_{[0,2 \pi)^d} \log \left\{ \det \Big( F \left( \Theta^{(d)}, u \right)  \Big) \right\} d \Theta^{(d)}_{unif} \right],
\end{align*}
where 
\begin{align*}
F \left( \wvec , u \right) = I_{2d} - u \widehat{M}_A (\wvec), 
\end{align*}
with $\wvec = (w_1, w_2, \ldots, w_d) \in \RM^d$.
\label{th001}
\end{theorem}

Furthermore, we introduce the {\em logarithmic zeta function} as follows. 
\begin{align*}
{\cal L} \left( A, T_{\infty}^d, u \right) = \log \left[ \lim_{N \to \infty} \left\{ \overline{\zeta} \left(A, T_N^d, u \right)^{-1} \right\} \right].
\end{align*}
This is closely related to the Mahler measure introduced in Section \ref{sec07m}. Then, the second equation in Theorem \ref{th001} immediately gives
\begin{theorem}
\begin{align*}
{\cal L} \left( A, T_{\infty}^d, u \right)
=
\int_{[0,2 \pi)^d} \log \left\{ \det \Big( I_{2d} - u \widehat{M}_A \left( \Theta^{(d)} \right) \Big) \right\} d \Theta^{(d)}_{unif}.
\end{align*}
\label{th001m}
\end{theorem}
\noindent
Note that throughout this paper, we assume ``$u \in \RM$" for our logarithmic zeta function ${\cal L} \left( A, T_{\infty}^d, u \right)$. The range of $u \in \RM$ depends on the model determined by a coin matrix $A$.

Moreover, we define $C_r (A, T^d_N)$ by
\begin{align}
\overline{\zeta} \left(A, T^d_N, u \right) = \exp \left( \sum_{r=1}^{\infty} \frac{C_r (A, T^d_N)}{r} u^r \right).
\label{satosan03}
\end{align}
Sometimes we write $C_r (A, T^d_N)$ by $C_r$ for short. Combining Eq. \eqref{satosan01} with Eq. \eqref{satosan03} implies
\begin{align*}
\det \Big( I_{2d N^d} - u M_A \Big)^{-1/N^d} = \exp \left( \sum_{r=1}^{\infty} \frac{C_r}{r} u^r \right).
\end{align*}
Thus we get
\begin{align}
- \frac{1}{N^d} \log \left\{ \det \Big( I_{2d N^d} - u M_A \Big) \right\} =  \sum_{r=1}^{\infty} \frac{C_r}{r} u^r.
\label{satosan05}
\end{align}
It follows from Eq. \eqref{keyresult01lemma} that the left-hand of Eq. \eqref{satosan05} becomes 
\begin{align*}
- \frac{1}{N^d} \log \left\{ \det \Big( I_{2d N^d} - u \widehat{M}_A \Big) \right\} 
&= 
- \frac{1}{N^d} \sum_{j=1}^{2d} \sum_{\kvec \in \mathbb{K}_N^d} \log \left\{ 1 - u \lambda_{j} (\kvec) \right\} 
\\
&
= \frac{1}{N^d} \sum_{j=1}^{2d} \sum_{\kvec \in \mathbb{K}_N^d} \sum_{r=1}^{\infty} \frac{\left(\lambda_{j} (\kvec) \right)^r}{r} u^r. 
\end{align*}
By this and the right-hand of Eq. \eqref{satosan05}, we have
\begin{align}
C_r (A, T^d_N) = \frac{1}{N^d} \sum_{j=1}^{2d} \sum_{\kvec \in \mathbb{K}_N^d} \left(\lambda_{j} (\kvec) \right)^r 
= \frac{1}{N^d} \sum_{j=1}^{2d} \sum_{\widetilde{\kvec} \in \widetilde{\mathbb{K}}_N^d} \left(\lambda_{j} (\widetilde{\kvec}) \right)^r.
\label{satosan06t}
\end{align}
Noting $\widetilde{k_j} = 2 \pi k_j/N \ (j=1,2, \ldots, d)$ and taking a limit as $N \to \infty$, we get
\begin{align}
\lim_{N \to \infty} C_r (A, T^d_N) = \sum_{j=1}^{2d} \int_{[0,2 \pi)^d} \lambda_{j} \left( \Theta^{(d)} \right)^r d \Theta^{(d)}_{unif}.
\label{satosan06limit}
\end{align}
Let ${\rm Tr} (A)$ denote the trace of a square matrix $A$. Then by definition of ${\rm Tr}$ and Eqs. \eqref{satosan06t} and \eqref{satosan06limit}, the following result was shown in \cite{K2}.

\begin{theorem}[Komatsu, Konno and Sato \cite{K2}]
\begin{align*}
C_r (A, T^d_N) 
&
= \frac{1}{N^d} \sum_{\widetilde{\kvec} \in \widetilde{\mathbb{K}}_N^d} {\rm Tr} \left( \left( \widehat{M}_A (\widetilde{\kvec}) \right)^r \right),
\nonumber
\\
\lim_{N \to \infty} C_r (A, T^d_N) 
&
= \int_{[0,2 \pi)^d} {\rm Tr} \left( \left( \widehat{M}_A (\Theta^{(d)}) \right)^r \right) d \Theta^{(d)}_{unif}
= {\rm Tr} \left( \Phi_r ^{(\infty)} ({\bf 0}) \right).
\end{align*}
\label{satosan06prop}
\end{theorem}
An interesting point is that $\Phi_r ^{(\infty)} ({\bf 0})$ is the return ``matrix weight" at time $r$ for the walk on not $T_N^d$ but $\ZM^d$. We should remark that in general ${\rm Tr} ( \Phi_r ^{(\infty)} ({\bf 0}) )$ is not the same as the return probability at time $r$ for QW and CRW, but for RW. 

Furthermore, we introduce
\begin{align*}
C_r (A, T^d_{\infty}) = \lim_{N \to \infty} C_r (A, T^d_N). 
\end{align*}
Therefore, by using the above equation, Theorem \ref{th001m}, and Eq. \eqref{satosan03}, we have
\begin{theorem}
\begin{align*}
{\cal L} \left( A, T_{\infty}^d, u \right)
= - \sum_{r=1}^{\infty} \frac{C_r (A, T^d_{\infty})}{r} \ u^r.
\end{align*}
\label{th002m}
\end{theorem}

From now on, we will present the result on only ${\cal L} \left( A, T_{\infty}^d, u \right)$ and $C_r (A, T^d_{\infty})$, since the corresponding expression for ``without $\lim_{N \to \infty}$" is the essentially same (see Theorems \ref{th001} and \ref{satosan06prop}, for example).

\section{Mahler Measure \label{sec07m}}
The {\it logarithmic Mahler measure} $m(f)$ of a nonzero $n$-variable Laurant polynomial $f(X_1, \ldots, X_n)$ is defined by
\begin{align}
m \left( f \right) = \int_{[0,1)^n} \log |f \left( e^{2 \pi i t_1}, \ldots, e^{2 \pi i t_n} \right) | \ d t_1 \cdots  d t_n.
\label{tonga01}
\end{align}
Note that 
\begin{align}
m \left( f \right) = \Re \left[ \int_{[0,1)^n} \log \left( f \left( e^{2 \pi i t_1}, \ldots, e^{2 \pi i t_n} \right) \right) \ d t_1 \cdots  d t_n \right],
\label{tonga01b}
\end{align}
where $\Re [z]$ is the real part of $z \in \CM$. Sometimes we simply refer to $m(f)$ as the {\it Mahler measure} of $f$. Then Eqs. \eqref{tonga01} and \eqref{tonga01b} can be rewritten as 
\begin{align}
m \left( f \right) 
&= \int_{[0,2 \pi)^n} \log |f \left( e^{i \theta_1}, \ldots, e^{i \theta_n} \right) | \ d \Theta^{(n)}_{unif}
\nonumber
\\
&= \Re \left[ \int_{[0,2 \pi)^n} \log \left( f \left( e^{i \theta_1}, \ldots, e^{i \theta_n} \right) \right) \ d \Theta^{(n)}_{unif} \right].
\label{tonga02}
\end{align}
This measure was introduced by Mahler \cite{M} in the study of number theory. As for Mahler measures, see \cite{Boyd1998, GR}, for example. We should remark that Akatsuka's zeta Mahler measure \cite{Akatsuka} is given by
\begin{align*}
Z \left( s, f \right) 
= \int_{[0,2 \pi)^n} |f \left( e^{i \theta_1}, \ldots, e^{i \theta_n} \right) |^s d \Theta^{(n)}_{unif},
\end{align*}
for a suitable $s \in \CM$. Thus it is different from our zeta function ${\cal L} \left( A, T_{\infty}^d, u \right)$ in Theorem \ref{th001m}: 
\begin{align*}
{\cal L} \left( A, T_{\infty}^d, u \right)
=
\int_{[0,2 \pi)^d} \log \left\{ \det \Big( I_{2d} - u \widehat{M}_A \left( \Theta^{(d)} \right) \Big) \right\} d \Theta^{(d)}_{unif}.
\end{align*}

For one-variable polynomials, it follows from the well-known Jensen formula that the following result (see \cite{Boyd1998, BM}, for example):
\begin{prop}
If
\begin{align*}
f(X) = a_0 \prod_{k=1}^n \left( X - \alpha_k \right),
\end{align*}
for $a_0 \in \CM \setminus \{0\}$ and $\alpha_k \in \CM$, then we have 
\begin{align*}
m (f) = \log |a_0| + \sum_{k=1}^n \log \left( \max \left\{ |\alpha_k|, 1 \right\} \right). 
\end{align*}
\label{Lemmaomikuron03}
\end{prop}
In particular, Proposition \ref{Lemmaomikuron03} gives
\begin{align*}
m \left( X + c \right) = \log \left( \max \{ |c|, 1 \} \right),
\end{align*}
for $c \in \CM$. Concerning multivariable polynomials, specific values of the Mahler measure involve the Dirichlet $L$-series and Riemann zeta functions. For example, the following formulas (Eqs. \eqref{tonga03} and \eqref{tonga05}) are due to Smyth \cite{Smyth}. 
\begin{align}
m \left( X_1 + X_2 + 1 \right)
= \frac{3 \sqrt{3}}{4 \pi} L \left( \chi_{-3}, 2 \right),
\label{tonga03}
\end{align}
where 
\begin{align*}
L \left( \chi_{-3}, s \right) = \sum_{n=1}^{\infty} \frac{\chi_{-3}(n)}{n^s} 
\end{align*}
is the Dirichlet $L$-series of the character
\begin{align*}
\chi_{-3}(n)
= \left\{ 
\begin{array}{cl}
1 & \mbox{if $n \equiv 1$ {\rm mod} 3,} 
\\
-1 & \mbox{if $n \equiv -1$ {\rm mod} 3,} 
\\
0 & \mbox{if $n \equiv 0$ {\rm mod} 3,}
\end{array}
\right.
\end{align*}
that is,
\begin{align*}
L \left( \chi_{-3}, 2 \right) = 1 - \frac{1}{2^2} + \frac{1}{4^2} - \frac{1}{5^2} + \cdots.
\end{align*}
Moreover, 
\begin{align}
m \left( X_1 + X_2 + X_3 + 1 \right)
= \frac{7}{2 \pi^2} \zeta (3),
\label{tonga05}
\end{align}
where $\zeta (s)$ is the Riemann zeta function given by
\begin{align*}
\zeta (s) = \sum_{n=1}^{\infty} \frac{1}{n^s} = \frac{1}{1^s} + \frac{1}{2^s} + \frac{1}{3^s} + \frac{1}{4^s} + \cdots.
\end{align*}

Now we present the following key result in order to compute the Mahler measure for one-dimensional QWs in Section \ref{sec03new}. 
\begin{lemma}
\begin{align}
m \left( X - X^{-1} + c \right)
&= \log \left( \frac{|c|+ \sqrt{c^2 + 4}}{2} \right)  \quad (c \in \RM),
\label{lemKeyM}
\\
m \left( X + X^{-1} + c \right)
&= \log \left( \frac{|c|+\sqrt{c^2 - 4}}{2} \right) \quad (c \in \RM \ \hbox{with} \ |c| \ge 2).
\label{lemKeyF}
\end{align}
\label{lemKeyLem}
\end{lemma}
\noindent
We should remark that Eq. \eqref{lemKeyM} is equivalent to
\begin{align*}
m \left( X - X^{-1} + c \right)
= \frac{1}{2} \log \left( \frac{c^2 + 2 + \sqrt{c^2 (c^2 + 4)}}{2} \right) \quad (c \in \RM).
\end{align*}
There are some proofs of Lemma \ref{lemKeyLem}. From now on, we will give two proofs, ``Proof A" and ``Proof B". The former is based on Proposition \ref{Lemmaomikuron03} and the latter comes from the following relation: for $r \in \RM \ \hbox{with} \ |r| \le 1$, 
\begin{align}
\int_0 ^{2 \pi} \log \left( 1 - r \cdot \sin \theta \right) \frac{d \theta}{2 \pi}
= 
\int_0 ^{2 \pi} \log \left( 1 - r \cdot \cos \theta \right) \frac{d \theta}{2 \pi}
= \log \left( \frac{1 + \sqrt{1 - r^2}}{2} \right).
\label{omikuron}
\end{align}
The first equality of Eq. \eqref{omikuron} is obvious. Concerning the second equality of Eq. \eqref{omikuron}, for example, it can be obtained by applying the Cauchy theorem and the fact that for $r \in (0,1)$,
\begin{align*}
\log \left( 1 - r \cdot \cos \theta \right) 
&= \log \left( \frac{r}{\sqrt{2}\sqrt{1- \sqrt{1- r^2}}} - \frac{\sqrt{1- \sqrt{1- r^2}}}{\sqrt{2}} \ e^{i \theta} \right)
\\
&+ \log \left( \frac{r}{\sqrt{2}\sqrt{1- \sqrt{1- r^2}}} - \frac{\sqrt{1- \sqrt{1- r^2}}}{\sqrt{2}} \ e^{-i \theta} \right).
\end{align*}
Note that it is enough to consider $r \in (0,1)$ for Eq. \eqref{omikuron}. In addition, other proofs of Eq. \eqref{lemKeyF} are found in, for example, Kurokawa and Ochiai \cite{KO}.
\par
\
\par 
\noindent
{\bf Proof A}. As for Eq. \eqref{lemKeyM}, we begin with
\begin{align*}
m \left( X - X^{-1} + c \right)
&= \int_0 ^{2 \pi} \log | e^{i \theta} - e^{-i \theta} + c| \ \frac{d \theta}{2 \pi}
\\
&= \int_0 ^{2 \pi} \log | e^{-i \theta} \left( e^{2 i \theta} - 1 + c e^{i \theta} \right) | \ \frac{d \theta}{2 \pi}
\\
&= \int_0 ^{2 \pi} \log | e^{2 i \theta} - 1 + c e^{i \theta}| \ \frac{d \theta}{2 \pi}
\\
&= 
m \left( X^2 + c X - 1 \right).
\end{align*}
Let 
\begin{align*}
f(X)=X^2 + c X - 1= \left( X - \alpha_1 \right) \left( X - \alpha_2 \right). 
\end{align*}
Then the solutions of $f(X)=0$ are given by
\begin{align*}
\alpha_1 = \frac{-c+\sqrt{c^2 + 4}}{2}, \qquad \alpha_2 = \frac{-c-\sqrt{c^2 + 4}}{2}.
\end{align*}
If $c \ge 0$, then we see that $\alpha_2 \le -1 < 0 < \alpha_1 \le 1$. So Proposition \ref{Lemmaomikuron03} implies 
\begin{align*}
m (f) 
&= \log \left( \max \left\{ |\alpha_1|, 1 \right\} \right) + \log \left( \max \left\{ |\alpha_2|, 1 \right\} \right)
\\
&= 0 + \log \left( \frac{c+\sqrt{c^2 + 4}}{2} \right).
\end{align*}
Thus we have
\begin{align}
m \left( X^2 + c X - 1 \right)
= \log \left( \frac{c+\sqrt{c^2 + 4}}{2} \right) \qquad (c \ge 0).
\label{umezu01}
\end{align}
Similarly, if $c \le 0$, then we have $- 1 \le \alpha_2 < 0 < 1 \le \alpha_1$. Therefore Proposition \ref{Lemmaomikuron03} gives
\begin{align*}
m (f) 
&= \log \left( \max \left\{ |\alpha_1|, 1 \right\} \right) + \log \left( \max \left\{ |\alpha_2|, 1 \right\} \right)
\\
&= \log \left( \frac{-c+\sqrt{c^2 + 4}}{2} \right) + 0.
\end{align*}
Thus we get
\begin{align}
m \left( X^2 + c X - 1 \right)
= \log \left( \frac{-c+\sqrt{c^2 + 4}}{2} \right) \qquad (c \le 0).
\label{umezu02}
\end{align}
Combining Eq. \eqref{umezu01} with Eq. \eqref{umezu02}, we obtain
\begin{align*}
m \left( X^2 + c X - 1 \right)
= \log \left( \frac{|c|+\sqrt{c^2 + 4}}{2} \right) \qquad (c \in \RM).
\end{align*}
Therefore the proof of Eq. \eqref{lemKeyM} is complete. In a similar way, as for Eq. \eqref{lemKeyF}, we start with
\begin{align*}
m \left( X + X^{-1} + c \right)
&= \int_0 ^{2 \pi} \log | e^{i \theta} + e^{-i \theta} + c| \ \frac{d \theta}{2 \pi}
\\
&= \int_0 ^{2 \pi} \log | e^{-i \theta} \left( e^{2 i \theta} + 1 + c e^{i \theta} \right) | \ \frac{d \theta}{2 \pi}
\\
&= \int_0 ^{2 \pi} \log | e^{2 i \theta} + 1 + c e^{i \theta}| \ \frac{d \theta}{2 \pi}
\\
&= 
m \left( X^2 + c X + 1 \right).
\end{align*}
Put 
\begin{align*}
f(X)=X^2 + c X + 1= \left( X - \alpha_1 \right) \left( X - \alpha_2 \right). 
\end{align*}
Then the solutions of $f(X)=0$ are given by
\begin{align*}
\alpha_1 = \frac{-c+\sqrt{c^2 - 4}}{2}, \qquad \alpha_2 = \frac{-c-\sqrt{c^2 - 4}}{2}.
\end{align*}
If $c \ge 2$, then we have $\alpha_2 \le -1 \le  \alpha_1 < 0$. So from Proposition \ref{Lemmaomikuron03}, we see
\begin{align*}
m (f) 
&= \log \left( \max \left\{ |\alpha_1|, 1 \right\} \right) + \log \left( \max \left\{ |\alpha_2|, 1 \right\} \right)
\\
&= 0 + \log \left( \frac{c+\sqrt{c^2 - 4}}{2} \right).
\end{align*}
Thus we get
\begin{align}
m \left( X^2 + c X + 1 \right)
= \log \left( \frac{c+\sqrt{c^2 - 4}}{2} \right) \qquad (c \ge 2).
\label{umezu01b}
\end{align}
Similarly, if $c \le -2$, then we have $ 0 < \alpha_2 \le 1 \le \alpha_1$. Therefore Proposition \ref{Lemmaomikuron03} implies
\begin{align*}
m (f) 
&= \log \left( \max \left\{ |\alpha_1|, 1 \right\} \right) + \log \left( \max \left\{ |\alpha_2|, 1 \right\} \right)
\\
&= \log \left( \frac{-c+\sqrt{c^2 - 4}}{2} \right) + 0.
\end{align*}
So we get
\begin{align}
m \left( X^2 + c X + 1 \right)
= \log \left( \frac{-c+\sqrt{c^2 - 4}}{2} \right) \qquad (c \le -2).
\label{umezu02b}
\end{align}
By Eqs. \eqref{umezu01b} and \eqref{umezu02b}, we have
\begin{align*}
m \left( X^2 + c X + 1 \right)
= \log \left( \frac{|c|+\sqrt{c^2 - 4}}{2} \right) \qquad (c \in \RM \ \hbox{with} \ |c| \ge 2).
\end{align*}
Thus the proof of Eq. \eqref{lemKeyF} is complete.
\hfill$\square$
\par
\
\par
From now on, we will give ``Proof B".
\par
\
\par
\noindent
{\bf Proof B}. Concerning Eq. \eqref{lemKeyM}, we see
\begin{align*}
m \left( X - X^{-1} + c \right)
&= \int_0 ^{2 \pi} \log | e^{i \theta} - e^{-i \theta} + c| \ \frac{d \theta}{2 \pi}
\\
&= \int_0 ^{2 \pi} \log | 2 i \sin \theta + c| \ \frac{d \theta}{2 \pi}
\\
&= \frac{1}{2} \int_0 ^{2 \pi} \log \left( 4 \sin^2 \theta + c^2 \right) \frac{d \theta}{2 \pi}
\\
&= \frac{1}{2} \int_0 ^{2 \pi} \log \left\{ c^2 + 2 - 2 \cos (2 \theta) \right\} \frac{d \theta}{2 \pi}
\\
&= \frac{1}{2} \log \left( c^2 + 2 \right) + \frac{1}{2} \int_0 ^{2 \pi} \log \left( 1 - \frac{2}{c^2 + 2} \cos \theta \right) \frac{d \theta}{2 \pi}
\\
&= \frac{1}{2} \log \left( \frac{c^2 + 2 + \sqrt{(c^2+2)^2 - 4}}{2} \right).
\end{align*}
The fifth equality can be derived from Eq. \eqref{omikuron}. Therefore, we obtain
\begin{align*}
m \left( X - X^{-1} + c \right)
&= \frac{1}{2} \log \left( \frac{c^2 + 2 + \sqrt{c^2 (c^2 + 4)}}{2} \right)
\\
&= \log \left[ \left( \frac{c^2 + 2 + |c| \sqrt{c^2 + 4}}{2} \right)^{1/2} \right]
\\
&= \log \left( \frac{|c| + \sqrt{c^2 + 4}}{2} \right).
\end{align*}
Thus the proof of Eq. \eqref{lemKeyM} is complete. Next, as for Eq. \eqref{lemKeyF}, we begin with
\begin{align*}
m \left( X + X^{-1} + c \right)
&= \int_0 ^{2 \pi} \log | e^{i \theta} + e^{-i \theta} + c| \ \frac{d \theta}{2 \pi}
\\
&= \int_0 ^{2 \pi} \log | 2 \cos \theta + c| \ \frac{d \theta}{2 \pi}.
\end{align*}
Noting that if $c \ge 2$, then ``$2 \cos \theta + c \ge 0$ for any $\theta \in [0, 2 \pi)$",  we see
\begin{align*}
m \left( X + X^{-1} + c \right)
&= \int_0 ^{2 \pi} \log \left( 2 \cos \theta + c \right) \ \frac{d \theta}{2 \pi}
\\
&= \log c + \int_0 ^{2 \pi} \log \left( 1  + \frac{2}{c} \cos \theta \right) \frac{d \theta}{2 \pi}
\\
&= \log c + \log \left( \frac{1 + \sqrt{1 - \left( \frac{2}{c} \right)^2}}{2} \right)
\\
&=\log \left( \frac{c+\sqrt{c^2 - 4}}{2} \right).
\end{align*}
In order to obtain the third equality, we used Eq. \eqref{omikuron} with $0 < 2/c \le 1$. In a similar fashion, noting that if $c \le -2$, then ``$2 \cos \theta + c \le 0$ for any $\theta \in [0, 2 \pi)$", we have
\begin{align*}
m \left( X + X^{-1} + c \right)
&= \int_0 ^{2 \pi} \log \left( - 2 \cos \theta - c \right) \ \frac{d \theta}{2 \pi}
\\
&= \log (-c) + \int_0 ^{2 \pi} \log \left( 1  + \frac{2}{c} \cos \theta \right) \frac{d \theta}{2 \pi}
\\
&= \log (-c) + \log \left( \frac{1 + \sqrt{1 - \left( \frac{2}{c} \right)^2}}{2} \right)
\\
&=\log \left( \frac{-c+\sqrt{c^2 - 4}}{2} \right).
\end{align*}
The third equality comes from Eq. \eqref{omikuron} with $-1 \le 2/c < 0$. Therefore the proof of Eq. \eqref{lemKeyF} is complete.
\hfill$\square$
\par
\
\par
By Lemma \ref{lemKeyLem}, we have the following result which will be used  for the one-dimensional QW with M-type and F-type in Section \ref{sec03new}.
\begin{lemma}
We assume that $\xi \in (0, \pi/2)$. 
\begin{align}
m \left( X - X^{-1} + c^{(m)} \right) = \log \left( \frac{ u - u^{-1} + \sqrt{u^2 + 2 \cos (2 \xi) + u^{-2}}}{2 \cos \xi} \right),
\label{lemKeyMM}
\end{align}
where $c^{(m)} = \sec \xi \cdot \left( u - u^{-1} \right)$ and $-1 < u < 0$.
\begin{align}
m \left( X + X^{-1} + c^{(f)} \right) = \log \left( \frac{ - \left(u + u^{-1} \right) + \sqrt{u^2 + 2 \cos (2 \xi) + u^{-2}}}{2 \sin \xi} \right),
\label{lemKeyFF}
\end{align}
where $c^{(f)} = - \ {\rm cosec} \ \xi \cdot \left( u + u^{-1} \right)$ and $ u < 0$.
\label{lemKeyLem2}
\end{lemma}
\noindent
{\bf Proof}. As for Eq. \eqref{lemKeyMM}, if we take $c= c^{(m)} = \sec \xi \cdot \left( u - u^{-1} \right)$ in Eq. \eqref{lemKeyM} of Lemma \ref{lemKeyLem}, then we have 
\begin{align*}
m \left( X - X^{-1} + c^{(m)} \right)
&= \log \left( \frac{\sec \xi \cdot \left( u - u^{-1} \right) + \sqrt{\sec^2 \xi \cdot \left(  u - u^{-1} \right)^2 +4}}{2} \right)
\\
&= \log \left( \frac{ u - u^{-1} + \sqrt{\left(  u - u^{-1} \right)^2 + 4 \cos^2 \xi}}{2 \cos \xi} \right)
\\
&= \log \left( \frac{ u - u^{-1} + \sqrt{u^2 + 2 \cos (2 \xi) + u^{-2}}}{2 \cos \xi} \right).
\end{align*}
The second equality comes from $\cos \xi >0$. In a similar fashion, as for Eq. \eqref{lemKeyFF}, if we take $c= c^{(f)} =  - \ {\rm cosec} \ \xi \cdot \left( u + u^{-1} \right)$ in Eq. \eqref{lemKeyF} of Lemma \ref{lemKeyLem}, then we have 
\begin{align*}
m \left( X + X^{-1} + c^{(f)} \right)
&= \log \left( \frac{ - {\rm cosec} \xi \cdot \left( u + u^{-1} \right) + \sqrt{{\rm cosec}^2 \xi \cdot \left(  u + u^{-1} \right)^2 - 4}}{2} \right)
\\
&= \log \left( \frac{ - \left(u + u^{-1} \right) + \sqrt{\left(  u + u^{-1} \right)^2 - 4 \sin^2 \xi}}{2 \sin \xi} \right)
\\
&= \log \left( \frac{ - \left(u + u^{-1} \right)+ \sqrt{u^2 + 2 \cos (2 \xi) + u^{-2}}}{2 \sin \xi} \right).
\end{align*}
\hfill$\square$

Here we present the result on the Mahler measure for two-variable case which will be used for the two-dimensional QW (in Section \ref{sec07}) and RW (in Section \ref{secmulti}). To do so, we introduce the {\em generalized hypergeometric function} which is defined by
\begin{align*}
{}_p F_q \left( a_1, \ldots, a_p ; b_1, \ldots, b_q ; x \right)
= \sum_{n=0}^{\infty} \frac{(a_1)_n \cdots (a_p)_n}{(b_1)_n \cdots (b_q)_n} \cdot \frac{x^n}{n!},
\end{align*}
where $(a)_n = \Gamma (a+n)/\Gamma (a)$ and $\Gamma (x)$ is the gamma function (see \cite{Andrews1999}, for example). Then the following result is given in Rodriguez-Villegas \cite{RV}.
\begin{lemma}
\begin{align*}
m \left( X_1 + X_1^{-1} + X_2 + X_2^{-1} + c \right)
=
\log c - \frac{2}{c^2} \ {}_4 F_3 \left( \frac{3}{2}, \frac{3}{2}, 1, 1 ; 2, 2, 2 ; \frac{16}{c^2} \right),
\end{align*}
where $c >4$.
\label{LemmaRV}
\end{lemma}

\section{One-Dimensional QW \label{sec03new}} 
In this section, we consider a relation between the logarithmic zeta function ${\cal L} \left( A, T_{\infty}^d, u \right)$ (introduced in Section \ref{sec02}) and the Mahler measure $m \left( f \right)$ (introduced in Section \ref{sec07m}) for two-state QWs on the one-dimensional torus $T_N^1$. As for a detailed study on the two-state QW, see \cite{K2, K6}.

Recall our setting introduced in Section \ref{sec02}. First we deal with general walks including QWs on the one-dimensional torus $T^1_N$ whose $2 \times 2$ coin matrix $A^{(m)}$ (M-type) or $A^{(f)}$ (F-type) as follows: 
\begin{align*}
A^{(m)}  
=
\begin{bmatrix}
a_{11} & a_{12} \\ 
a_{21} & a_{22} 
\end{bmatrix} 
, \qquad 
A^{(f)} 
=  
\begin{bmatrix}
a_{21} & a_{22} \\ 
a_{11} & a_{12} 
\end{bmatrix}
,
\end{align*}
since
\begin{align*}
A^{(f)} = \left( I_1 \otimes \sigma \right) A^{(m)} = \sigma  A^{(m)} = 
\begin{bmatrix}
0 & 1 \\ 
1 & 0 
\end{bmatrix}
\begin{bmatrix}
a_{11} & a_{12} \\ 
a_{21} & a_{22} 
\end{bmatrix} 
.
\end{align*}
Set $k=k_1$ and $\widetilde{k} = \widetilde{k}_1$. In this case, we take 
\begin{align*}
P_{1} 
=
\begin{bmatrix}
1 & 0 \\ 
0 & 0 
\end{bmatrix}
, \qquad 
P_{2} 
=
\begin{bmatrix}
0 & 0 \\ 
0 & 1 
\end{bmatrix}
. 
\end{align*}
Thus we immediately get
\begin{align*} 
\widehat{M}_{A^{(m)}} (\widetilde{k})
&= e^{i \widetilde{k}} P_{1} A^{(m)} + e^{-i \widetilde{k}} P_{2} A^{(m)} 
= 
\begin{bmatrix}
e^{i \widetilde{k}} a_{11} & e^{i \widetilde{k}} a_{12} \\ 
e^{-i \widetilde{k}}a_{21} & e^{-i \widetilde{k}} a_{22} 
\end{bmatrix} 
,
\\
\widehat{M}_{A^{(f)}} (\widetilde{k})
&= e^{i \widetilde{k}} P_{1} A^{(f)} + e^{-i \widetilde{k}} P_{2} A^{(f)} 
= 
\begin{bmatrix}
e^{i \widetilde{k}} a_{21} & e^{i \widetilde{k}} a_{22} \\ 
e^{-i \widetilde{k}}a_{11} & e^{-i \widetilde{k}} a_{12} 
\end{bmatrix} 
.
\end{align*}
By these equations, we have
\begin{align*}
\det \Big( I_{2} - u \widehat{M}_{A^{(s)}} (\widetilde{k}) \Big) 
= 1 - {\rm Tr} \left(  \widehat{M}_{A^{(s)}} (\widetilde{k}) \right) u + \det \left(  \widehat{M}_{A^{(s)}} (\widetilde{k}) \right) u^2 \qquad (s \in \{m,f\}).
\end{align*}
Then the result given in \cite{K2} can be rewritten in terms of the logarithmic zeta function ${\cal L} \left( A, T_{\infty}^d, u \right)$ as follows:
\begin{prop}
\begin{align*}
{\cal L} \left( A^{(s)}, T_{\infty}^1, u \right) 
= \int_0^{2 \pi} \log \left\{ 1 - {\rm Tr} \left(  \widehat{M}_{A^{(s)}}  (\theta) \right) u + \det \left(  \widehat{M}_{A^{(s)}} (\theta) \right) u^2 \right\} \frac{d \theta}{2 \pi},
\end{align*}
for $s \in \{m,f\}$. 
\label{kimarid1}
\end{prop}
\noindent
Remark that Proposition \ref{kimarid1} is also obtained by Theorem \ref{th001m}.

From now on, we focus on QWs in one dimension. One of the typical classes of QWs for $2 \times 2$ coin matrix $A^{(m)}$ (M-type) or $A^{(f)}$ (F-type) is as follows: 
\begin{align*}
A^{(m)} 
=
\begin{bmatrix}
\cos \xi & \sin \xi  \\ 
\sin \xi & - \cos \xi  
\end{bmatrix} 
, \qquad 
A^{(f)} 
=  
\begin{bmatrix}
\sin \xi & - \cos \xi \\ 
\cos \xi & \sin \xi  
\end{bmatrix}
\qquad ( \xi \in [0, 2\pi)).
\end{align*}
When $\xi = \pi/4$, the QW becomes the so-called {\em Hadamard walk} which is one of the most well-investigated model in the study of QWs. Then the result given in \cite{K2} can also be rewritten in terms of the logarithmic zeta function like Proposition \ref{kimarid1} as follows:
\begin{prop} 
\begin{align*}
{\cal L} \left( A^{(s)}, T_{\infty}^1, u \right) 
= \int_0^{2 \pi} \log \left( F^{(s)} \left( \theta, u \right) \right) \frac{d \theta}{2 \pi},
\end{align*}
for $s \in \{m,f\}$, where
\begin{align*}
F^{(m)} \left( w, u \right) 
&= 1 - 2 i \cos \xi \sin w \cdot u - u^2,
\\
F^{(f)} \left( w, u \right) 
&= 1 - 2 \sin \xi \cos w \cdot u + u^2.
\end{align*}
\label{kimarid1qw}
Moreover, 
\begin{align*}
C_{2l} (A^{(m)}, T^1_{\infty}) 
&= 2 l \left(-  \cos^2 \xi \right)^{l} \sum_{m=1}^l \frac{1}{m} {l-1 \choose m-1}^2 \left( - \tan^2 \xi \right)^{m}
\\
&= 2 l \left(-  \cos^2 \xi \right)^{l-1} (\sin^2 \xi) \ {}_2F_1 \left( 1-l , 1-l ; 2 ; - \tan^2 \xi \right),
\\
C_{2l} (A^{(f)}, T^1_{\infty}) 
&= 2 l \left(\sin \xi \right)^{2l} \sum_{m=1}^l \frac{1}{m} {l-1 \choose m-1}^2 \left( - \cot^2 \xi \right)^{m}
\\
&= 2 l \left(\sin \xi \right)^{2(l-1)} (- \cos^2 \xi) \ {}_2F_1 \left( 1-l , 1-l ; 2 ; - \cot^2 \xi \right),
\\
C_{2l-1} (A^{(s)}, T^1_{\infty}) 
&= 0 \qquad (s \in \{m,f\}), 
\end{align*}
for $l=1,2, \ldots$ and $\xi \in (0,\pi/2).$ 
\label{kimarid1qwcr}
\end{prop}
Note that the result on ${\cal L} \left( A^{(s)}, T_{\infty}^1, u \right)$ in Proposition \ref{kimarid1qwcr} is also derived from Proposition \ref{kimarid1}. Here we present one of our main results: 
\begin{theorem}
Let $c^{(m)} = \sec \xi \cdot \left( u - u^{-1} \right)$ and $c^{(f)} = - \ {\rm cosec} \ \xi \cdot \left( u + u^{-1} \right)$. Then we have
\begin{align}
{\cal L} \left( A^{(m)}, T_{\infty}^1, u \right) 
&= \log \left( \frac{ 1 - u^{2} + \sqrt{1 + 2 \cos (2 \xi) u^2 + u^{4}}}{2} \right)
\label{senreim01}
\\
&= \log \left( - \cos \xi \cdot u \right) + m \left( X - X^{-1} + c^{(m)} \right),
\label{senreim02}
\end{align}
for $\xi \in (0, \pi/2)$ and $u \in (\cos \xi - \sqrt{\cos^2 \xi +1}, 0)$.
\begin{align}
{\cal L} \left( A^{(f)}, T_{\infty}^1, u \right) 
&= \log \left( \frac{ 1 + u^{2} + \sqrt{1 + 2 \cos (2 \xi) u^2 + u^{4}}}{2} \right)
\label{senreif01}
\\
&= \log \left( - \sin \xi \cdot u \right) + m \left( X + X^{-1} + c^{(f)} \right),
\label{senreif02}
\end{align}
for $\xi \in (0, \pi/2)$ and $u \in (- \infty, 0)$. Here
\begin{align}
m \left( X - X^{-1} + c^{(m)} \right) 
&= \log \left( \frac{ u - u^{-1} + \sqrt{u^2 + 2 \cos (2 \xi) + u^{-2}}}{2 \cos \xi} \right),
\label{senreim03}
\\
m \left( X + X^{-1} + c^{(f)} \right) 
&= \log \left( \frac{ - \left(u + u^{-1} \right) + \sqrt{u^2 + 2 \cos (2 \xi) + u^{-2}}}{2 \sin \xi} \right).
\label{senreif03}
\end{align}
\label{mainth05}
\end{theorem}
\noindent
{\bf Proof}.
As for Eq. \eqref{senreim01}, we start with
\begin{align}
{\cal L} \left( A^{(m)}, T_{\infty}^1, u \right) 
&= 
\int_0^{2 \pi} \log \left( 1 - 2 i \cos \xi \sin \theta \cdot u - u^2 \right) \frac{d \theta}{2 \pi}
\nonumber
\\
&= 
\log \left( 1 - u^{2} \right)  
+ \int_0^{2 \pi} \log \left( 1 - \frac{2 i \cos \xi \cdot u}{1 - u^2} \sin \theta \right) \frac{d \theta}{2 \pi}.
\label{kazuom01}
\end{align}
The first equality is given by Proposition \ref{kimarid1qwcr}. In order to get the second equality, we used $1 - u^2 > 0$, since  $- 1 < (\cos \xi - \sqrt{\cos^2 \xi +1} \le) u < 0$ for $\xi \in (0, \pi/2)$. Note that if $- 1 < \cos \xi - \sqrt{\cos^2 \xi +1} \le u < 0$, then 
\begin{align*}
\left| \frac{2 i \cos \xi \cdot u}{1 - u^2} \right| \le 1.
\end{align*}
Therefore combining Eq. \eqref{kazuom01} with Eq. \eqref{omikuron} implies 
\begin{align*}
{\cal L} \left( A^{(m)}, T_{\infty}^1, u \right) 
&= \log \left( 1 - u^{2} \right)  
+ \log \left( \frac{ 1 + \sqrt{1 - \left( \frac{2 i \cos \xi \cdot u}{1 - u^2} \right)^2}}{2} \right)
\\
&= \log \left( \frac{ 1 - u^2 + \sqrt{(1-u^2)^2 + 4 \cos ^2 \xi \cdot u^2}}{2} \right)
\\
&= \log \left( \frac{ 1 - u^{2} + \sqrt{1 + 2 \cos (2 \xi) u^2 + u^{4}}}{2} \right),
\end{align*}
for $\xi \in (0, \pi/2)$ and $u \in (\cos \xi - \sqrt{\cos^2 \xi +1}, 0)$. Remark that as for Eq. \eqref{omikuron}, we can also use ``$i r$" instead of ``$r$" by a standard argument of complex analysis. So the proof of Eq. \eqref{senreim01} is complete.

On the other hand, Eq. \eqref{senreim03} is equivalent to Eq. \eqref{lemKeyMM} in Lemma \ref{lemKeyLem2}. Moreover, we see that $- \cos \xi \cdot u >0$, since $\xi \in (0, \pi/2)$ and $u \in (-1, 0)$. Thus we have 
\begin{align*}
&\log \left( - \cos \xi \cdot u \right) + m \left( X - X^{-1} + c^{(m)} \right),\\
&=\log \left( - \cos \xi \cdot u \right) + \log \left( \frac{ u - u^{-1} + \sqrt{u^2 + 2 \cos (2 \xi) + u^{-2}}}{2 \cos \xi} \right)
\\
&= \log \left( \frac{ 1 - u^{2} + \sqrt{1 + 2 \cos (2 \xi) u^2 + u^{4}}}{2} \right).
\end{align*}
The first equality is given by Eq. \eqref{senreim03}. So the proof of  Eq. \eqref{senreim02} is complete.

Next we move to the proof of Eq. \eqref{senreif01}. We begin with
\begin{align}
{\cal L} \left( A^{(f)}, T_{\infty}^1, u \right) 
&= 
\int_0^{2 \pi} \log \left( 1 - 2 \sin \xi \cos \theta \cdot u + u^2 \right) \frac{d \theta}{2 \pi}
\nonumber
\\
&= 
\log \left( 1 + u^{2} \right)  
+ \int_0^{2 \pi} \log \left( 1 - \frac{2 \sin \xi \cdot u}{1 + u^2} \cos \theta \right) \frac{d \theta}{2 \pi}.
\label{kazuof01}
\end{align}
The first equality is given by Proposition \ref{kimarid1qwcr}. The second equality comes from $1+u^2 > 0$ for any $u \in \RM$. Note that 
\begin{align*}
\left| \frac{2 \sin \xi \cdot u}{1 + u^2} \right| \le 1,
\end{align*}
for any $u \in \RM$. Therefore combining Eq. \eqref{kazuof01} with Eq. \eqref{omikuron} implies 
\begin{align*}
{\cal L} \left( A^{(f)}, T_{\infty}^1, u \right) 
&= \log \left( 1 + u^{2} \right)  
+ \log \left( \frac{ 1 + \sqrt{1 - \left( \frac{2 \sin \xi \cdot u}{1 + u^2} \right)^2}}{2} \right)
\\
&= \log \left( \frac{ 1 + u^2 + \sqrt{(1+u^2)^2 - 4 \sin ^2 \xi \cdot u^2}}{2} \right)
\\
&= \log \left( \frac{ 1 + u^{2} + \sqrt{1 + 2 \cos (2 \xi) u^2 + u^{4}}}{2} \right),
\end{align*}
for $\xi \in (0, \pi/2)$ and $u \in \RM$. So the proof of Eq. \eqref{senreif01} is complete.

On the other hand, Eq. \eqref{senreif03} is equivalent to Eq. \eqref{lemKeyFF} in Lemma \ref{lemKeyLem2}. Furthermore, we see that $- \sin \xi \cdot u >0$, since $\xi \in (0, \pi/2)$ and $u \in (- \infty, 0)$. Thus we get
\begin{align*}
&\log \left( - \sin \xi \cdot u \right) + m \left( X + X^{-1} + c^{(f)} \right),\\
&=\log \left( - \sin \xi \cdot u \right) + \log \left( \frac{ - \left( u + u^{-1} \right) + \sqrt{u^2 + 2 \cos (2 \xi) + u^{-2}}}{2 \sin \xi} \right)
\\
&= \log \left( \frac{ 1 + u^{2} + \sqrt{1 + 2 \cos (2 \xi) u^2 + u^{4}}}{2} \right).
\end{align*}
The first equality is given by Eq. \eqref{senreif03}. So the proof of  Eq. \eqref{senreif02} is complete.
\hfill$\square$
\par
\
\par
In particular, when $\xi = \pi/4$ (Hadamard walk), Theorem \ref{mainth05} gives
\begin{cor}
\begin{align*}
{\cal L} \left( A^{(m)}, T_{\infty}^1, u \right) 
&= \log \left( \frac{1-u^2 + \sqrt{1+u^4}}{2} \right)
\\
&= \log \left( - \frac{u}{\sqrt{2}} \right) + m \left( X - X^{-1} + \sqrt{2} (u - u^{-1}) \right), 
\end{align*}
for $u \in ((\sqrt{2}-\sqrt{6})/2, 0)$, where
\begin{align*}
m \left( X - X^{-1} + \sqrt{2} (u - u^{-1}) \right) = \log \left( \frac{u-u^{-1} + \sqrt{u^2+u^{-2}}}{\sqrt{2}} \right).
\end{align*}
\begin{align*}
{\cal L} \left( A^{(f)}, T_{\infty}^1, u \right) 
&= \log \left( \frac{1+u^2 + \sqrt{1+u^4}}{2} \right)
\\
&= \log \left( - \frac{u}{\sqrt{2}} \right) + m \left( X + X^{-1} - \sqrt{2} (u + u^{-1}) \right),
\end{align*}
for $u \in (- \infty, 0)$, where
\begin{align*}
m \left( X + X^{-1} - \sqrt{2} (u + u^{-1}) \right) = \log \left( \frac{- \left(u+u^{-1} \right) + \sqrt{u^2+u^{-2}}}{\sqrt{2}} \right).
\end{align*}
\end{cor}

We should remark that in the case of the symmetric RW, $2 \times 2$ coin matrices $A^{(m)}$ (M-type) and $A^{(f)}$ (F-type) become 
\begin{align*}
A^{(m)} 
= A^{(f)} = \frac{1}{2}
\begin{bmatrix}
1 & 1  \\ 
1 & 1  
\end{bmatrix} 
.
\end{align*}
Then we have
\begin{cor}
\begin{align*}
{\cal L} \left( A^{(s)}, T_{\infty}^1, u \right)
&= \int_0^{2 \pi} \log \left( 1 -  u \cdot \cos \theta \right) \frac{d \theta}{2 \pi} = \log \left( \frac{1 + \sqrt{1 - u^2}}{2} \right)
\\
&= \log ( - u/2) + m \left( X + X^{-1} - 2 u^{-1} \right) \quad ( -1<u<0), 
\end{align*}
for $s \in \{m,f\}$ and $l=1,2, \ldots$.
\label{kimarid1symrw}
\end{cor}
\noindent
The first equality is given by Proposition \ref{kimarid1}. Remark that $\det \left(  \widehat{M}_{A^{(s)}} (\theta) \right) = 0$ implies that the coefficient of $u^2$ is $0$. The second equality comes from Eq. \eqref{omikuron}. The proof of the third equality can be found in the proof of Proposition \ref{mitakeumi00}.  In Section \ref{secmulti}, we will deal with the case of $d$-dimensional RWs.

\section{Higher-Dimensional QW \label{sec07}} 
In this section, we consider the $d$-dimensional QW, in particular, the Grover walk which is one of the well-investigated model in the study of the QW. The $2d \times 2d$ coin matrix of the $d$-dimensional {\em Grover walk} is defined by the following  Grover matrix $A =[A_{ij}] $:
\begin{align*}
A_{ij}= \frac{1}{d} - \delta_{ij}
=
\begin{cases}
\frac{1}{d} - 1 & (i=j), \\
\frac{1}{d} & (i \neq j),
\end{cases}
\end{align*}
where $\delta_{ij} = 1 \ (i=j), = 0 \ (i \not= j)$. Then $A$ is unitary. 
For example, when $d=2$, we treat the four-state model on the two-dimensional torus $T_N^2$ defined by the following $4 \times 4$ coin matrix $A^{(m)}$ (M-type) and $A^{(f)}$ (F-type): 
\begin{align*}
A^{(m)} &=
\begin{bmatrix}
-\frac{1}{2} & \frac{1}{2} & \frac{1}{2} & \frac{1}{2} \\
\frac{1}{2} & -\frac{1}{2} & \frac{1}{2} & \frac{1}{2} \\
\frac{1}{2} & \frac{1}{2} & -\frac{1}{2} & \frac{1}{2} \\
\frac{1}{2} & \frac{1}{2} & \frac{1}{2} & -\frac{1}{2} 
\end{bmatrix}, 
\\
A^{(f)} &=
\begin{bmatrix}
\frac{1}{2} & -\frac{1}{2} & \frac{1}{2} & \frac{1}{2} \\
-\frac{1}{2} & \frac{1}{2} & \frac{1}{2} & \frac{1}{2} \\
\frac{1}{2} & \frac{1}{2} & \frac{1}{2} & -\frac{1}{2} \\
\frac{1}{2} & \frac{1}{2} & -\frac{1}{2} & \frac{1}{2} 
\end{bmatrix}
= \left(I_{2} \otimes \sigma \right) A^{(m)}. 
\end{align*}
In order to consider the $d$-dimensional case, we focus on F-type, since the corresponding result for M-type is limited (see \cite{K2}). The result of Corollary 1 in \cite{K1} (Grover/Zeta Correspondence), or equivalently, Corollary 14 in \cite{K2} (Walk/Zeta Correspondence) can be rewritten in terms of the logarithmic zeta function as follows:
\begin{prop}
\begin{align*}
{\cal L} \left( A^{(f)}, T_{\infty}^d, u \right)
&=(d-1) \log (1-u^2) 
+ \int_{[0, 2 \pi)^d} \log\left\{F^{(f)}(\Theta^{(d)}, u)\right\} d \Theta^{(d)}_{unif},
\end{align*}
for $u \in (-1, 1)$, where
\begin{align*}
F^{(f)}(\wvec, u) 
=1 - \frac{2 \ e^{(d, \cos)}(\wvec)}{d} \cdot u + u^2.
\end{align*}
Here $\Theta^{(d)} = (\theta_1, \theta_2, \ldots, \theta_d) \in [0, 2 \pi)^d, \ \wvec = (w_1, w_2, \ldots, w_d) \in \RM^d$, 
\begin{align*}
e^{(d, \cos)} (\wvec) = \sum_{j=1}^d \cos w_j, 
\end{align*}
and $d \Theta^{(d)}_{unif}$ denotes the uniform measure on $[0, 2 \pi)^d$, that is,
\begin{align*}
d \Theta^{(d)}_{unif} = \frac{d \theta_1}{2 \pi } \cdots \frac{d \theta_d}{2 \pi }.
\end{align*}
\label{koredayo01}
\end{prop}
\noindent
Furthermore, we have the following one of the main results which gives a relationship between our logarithmic zeta function and the Mahler measure for the $d$-dimensional Grover walk with F-type.
\begin{theorem}
\begin{align*}
{\cal L} \left( A^{(f)}, T_{\infty}^d, u \right)
=(d-1) \log (1-u^2) + \log \left( - \frac{u}{d} \right) + m \left( \sum_{j=1}^d \left( X_j + X_j^{-1} \right) + c \right),
\end{align*}
for $u \in (-1, 0)$, where $c = - d \left( u +  u^{-1} \right)$.
\label{koredayo01m}
\end{theorem}
\noindent
{\bf Proof}. By Proposition \ref{koredayo01}, we start with
\begin{align*}
&{\cal L} \left( A^{(f)}, T_{\infty}^d, u \right)
\\
& \qquad =(d-1) \log (1-u^2) 
+ \int_{[0, 2 \pi)^d} \log\left\{ 1 - \frac{2}{d} \left( \sum_{j=1}^d \cos \theta_j \right) \cdot u + u^2 \right\} d \Theta^{(d)}_{unif}
\\
& \qquad =(d-1) \log (1-u^2) 
+ \int_{[0, 2 \pi)^d} \log\left\{ 1 - \frac{u}{d} \sum_{j=1}^d \left( e^{i \theta_j} + e^{-i \theta_j} \right) + u^2 \right\} d \Theta^{(d)}_{unif}
\\
& \qquad =(d-1) \log (1-u^2) + \log \left( - \frac{u}{d} \right) 
\\
& \qquad \qquad \qquad + \int_{[0, 2 \pi)^d} \log\left\{ \sum_{j=1}^d \left( e^{i \theta_j} + e^{-i \theta_j} \right) - d \left( u +  u^{-1} \right) \right\} d \Theta^{(d)}_{unif}.
\end{align*}
We see that if $u \in (-1,0)$, then 
\begin{align*}
\sum_{j=1}^d \left( e^{i \theta_j} + e^{-i \theta_j} \right) - d \left( u +  u^{-1} \right) > 0.
\end{align*}
Noting this and Eq. \eqref{tonga02}, we get 
\begin{align*}
&\int_{[0, 2 \pi)^d} \log\left\{ \sum_{j=1}^d \left( e^{i \theta_j} + e^{-i \theta_j} \right) - d \left( u +  u^{-1} \right) \right\} d \Theta^{(d)}_{unif}
\\
& \qquad = \Re \left[ \int_{[0, 2 \pi)^d} \log\left\{ \sum_{j=1}^d \left( e^{i \theta_j} + e^{-i \theta_j} \right) - d \left( u +  u^{-1} \right) \right\} d \Theta^{(d)}_{unif} \right]
\\
&\qquad = m \left( \sum_{j=1}^d \left( X_j + X_j^{-1} \right) - d \left( u +  u^{-1} \right) \right).
\end{align*}
Therefore we have the desired conclusion.
\hfill$\square$
\par
\
\par
When $d=2$, it follows from Theorem \ref{koredayo01m} that 
\begin{align}
{\cal L} \left( A^{(f)}, T_{\infty}^2, u \right)
= \log (1-u^2) +  \log \left( - \frac{u}{2} \right) + m \left( \sum_{j=1}^2 \left( X_j + X_j^{-1} \right) - 2 \left( u +  u^{-1} \right) \right).
\label{omikuron01c}
\end{align}
As for the third term in the right-hand side of Eq. \eqref{omikuron01c}, if we take $c= - 2 \left( u +  u^{-1} \right)$ with $u \in (-1,0)$ in Lemma \ref{LemmaRV}, then we have
\begin{align}
m \left( \sum_{j=1}^2 \left( X_j + X_j^{-1} \right) - 2 \left( u +  u^{-1} \right) \right)
=\log \left( - 2 \left( u +  u^{-1} \right) \right) - \frac{2}{c^2} \ {}_4 F_3 \left( \frac{3}{2}, \frac{3}{2}, 1, 1 ; 2, 2, 2 ; \frac{16}{c^2} \right).
\label{omikuron02c}
\end{align}
Combining Eq. \eqref{omikuron01c} with Eq. \eqref{omikuron02c} implies
\begin{align*}
{\cal L} \left( A^{(f)}, T_{\infty}^2, u \right)
&= \log \left( 1-u^2 \right) + \log \left( - \frac{u}{2} \right) + \log \left(  - 2 \left( u +  u^{-1} \right) \right) 
\\
& \qquad \qquad \qquad \qquad - \frac{2}{c^2} \ {}_4 F_3 \left( \frac{3}{2}, \frac{3}{2}, 1, 1 ; 2, 2, 2 ; \frac{16}{c^2} \right)
\\
&= \log \left\{ (1-u^2)  (1+u^2) \right\}  
- \frac{2}{c^2} \ {}_4 F_3 \left( \frac{3}{2}, \frac{3}{2}, 1, 1 ; 2, 2, 2 ; \frac{16}{c^2} \right).
\end{align*}
Therefore we obtain
\begin{cor}
\begin{align*}
{\cal L} \left( A^{(f)}, T_{\infty}^2, u \right)
= \log \left( 1-u^4  \right) 
- \frac{2}{c^2} \ {}_4 F_3 \left( \frac{3}{2}, \frac{3}{2}, 1, 1 ; 2, 2, 2 ; \frac{16}{c^2} \right),
\end{align*}
for $u \in (-1, 0)$, where $c = - 2 (u + u^{-1})$.
\end{cor}

\section{Higher-Dimensional RW \label{secmulti}}
For comparison with QW, we deal with $d$-dimensional RWs. As for detailed information on RWs in higher dimensions related to this section, see \cite{K5}. 

Let 
\begin{align*}
e^{(n)} (\xvec) = e^{(n)} (x_1, x_2, \ldots, x_n) = x_1 + x_2 + \cdots + x_n,
\end{align*}
for $\xvec = (x_1, x_2, \ldots, x_n) \in \CM^n$ and $n \in \ZM_{>}.$ Recall that $\mathbb{K}_N = \{ 0,1, \ldots, N-1 \}$ and $\widetilde{\mathbb{K}}_N = \{ 0 ,2 \pi/N, \ldots, 2 \pi (N-1)/N \}$. For $\kvec=(k_1,k_2,\ldots,k_{d}) \in \mathbb{K}_N^d$, we define 
\begin{align*}
\widetilde{k}_j = \frac{2 \pi k_j}{N} \in \widetilde{\mathbb{K}}_N, \quad \widetilde{\kvec}=(\widetilde{k}_1,\widetilde{k}_2,\ldots,\widetilde{k}_{d}) \in \widetilde{\mathbb{K}}_N^d.
\end{align*}
In this setting, we introduce
\begin{align*}
e^{(n, \cos)} (\widetilde{\kvec}) 
=  e^{(n)} (\cos \widetilde{k}_1, \cos \widetilde{k}_2, \ldots, \cos \widetilde{k}_n) \qquad (n \in \ZM_{>}).
\end{align*}
Moreover, for $\Theta^{(n)} = (\theta_1, \theta_2, \ldots, \theta_n) \in [0, 2 \pi)^n$,  
\begin{align*}
e^{(n, \cos)} (\Theta^{(n)}) =  e^{(n)} (\cos \theta_1, \cos \theta_2, \ldots, \cos \theta_n) \qquad (n \in \ZM_{>}).
\end{align*}
Note that, for $\wvec = (w_1, w_2, \ldots, w_n) \in \RM^n$, 
\begin{align*}
e^{(n, \cos)} (\wvec) 
=  e^{(n)} (\cos w_1, \cos w_2, \ldots, \cos w_n) \qquad (n \in \ZM_{>}).
\end{align*}

From definition of the (simple symmetric) RW on $T^d_N$ (see \cite{Norris, Spitzer}), we easily see that 
\begin{align} 
{\rm Spec} \left( P^{(D,c)} \right) 
&= \left\{ \frac{1}{d} \sum^d_{j=1} \cos \left( \frac{2 \pi k_j }{N} \right) \bigg| \ k_1 , \ldots , k_d \in \mathbb{K}_N \right\} 
\nonumber
\\
&= \left\{ \frac{1}{d} e^{(d, \cos)} (\widetilde{\kvec}) \bigg| \ k_1 , \ldots , k_d \in \mathbb{K}_N \right\},
\label{specP}
\end{align}
where $P^{(D,c)}$ is the transition probability matrix of the (simple symmetric) RW on $T^d_N$. Here the RW on $T^d_N$ jumps to each of its nearest neighbors with equal probability $1/(2d)$. Noting Eq. \eqref{specP}, the result given in \cite{K5} can be rewritten in terms of the logarithmic zeta function as follows:
\begin{prop}
\begin{align*}
{\cal L} \left( A_{RW}, T_{\infty}^d, u \right)
= \int_{[0,2 \pi)^d} \log \Bigg\{ F_{RW} \left( \Theta^{(d)}, u \right) \Bigg\} d \Theta^{(d)}_{unif},
\end{align*}
where 
\begin{align*}
F_{RW} \left( \wvec, u \right) = 1 -\frac{e^{(d, \cos)}(\wvec)}{d} \cdot u.
\end{align*}
Moreover, we have
\begin{align*}
C_{r} (A_{RW},T_{\infty}^d)
&= \int_{[0,2 \pi)^d} G_{RW} \left( \Theta^{(d)} \right) d \Theta^{(d)}_{unif}, 
\end{align*}
where 
\begin{align*}
G_{RW} \left( \wvec \right) = \left( \frac{e^{(d, \cos)} (\wvec)}{d} \right)^r.
\end{align*} 
\label{prop002D}
\end{prop}
From Proposition \ref{prop002D}, we obtain the following theorem for RW which corresponds to Theorem \ref{koredayo01m} for QW.
\begin{theorem}
\begin{align}
{\cal L} \left( A_{RW}, T_{\infty}^d, u \right)
= \log \left( - \frac{u}{2d} \right) + m \left( \sum_{j=1}^{d} (X_j + X_j^{-1}) - \frac{2d}{u} \right) \quad (-1<u<0).
\label{ikagame01}
\end{align}
\label{abi01}
\end{theorem}
\noindent
{\bf Proof}. As in the proof of Theorem \ref{koredayo01m}, we begin with
\begin{align*}
&{\cal L} \left( A_{RW}, T_{\infty}^d, u \right)
\\
& = \int_{[0, 2 \pi)^d} \log \left( 1 - \frac{u}{d} \cdot \sum_{j=1}^{d} \cos \theta_j \right) d \Theta^{(d)}_{unif}
\\
& = \int_{[0, 2 \pi)^d} \log \left\{ 1 -  \frac{u}{2d} \cdot \sum_{j=1}^{d} \left( e^{i \theta_j} +  e^{-i \theta_j} \right) \right\} d \Theta^{(d)}_{unif}
\\
& = \log \left( - \frac{u}{2d} \right) 
+ \int_{[0, 2 \pi)^d} \log \left\{ \sum_{j=1}^{d} \left( e^{i \theta_j} +  e^{-i \theta_j} \right) - \frac{2d}{u} \right\} d \Theta^{(d)}_{unif}.
\end{align*}
On the other hand, we see that if $u \in (-1,0)$, then 
\begin{align}
\sum_{j=1}^{d} \left( e^{i \theta_j} +  e^{-i \theta_j} \right) - \frac{2d}{u} > 0. 
\label{tonga001}
\end{align}
Noting Eqs. \eqref{tonga02} and \eqref{tonga001}, we obtain
\begin{align*}
&\int_{[0, 2 \pi)^d} \log \left\{ \sum_{j=1}^{d} \left( e^{i \theta_j} +  e^{-i \theta_j} \right) - \frac{2d}{u} \right\} d \Theta^{(d)}_{unif}
\\
&= \Re \left[ \int_{[0, 2 \pi)^d} \log \left\{ \sum_{j=1}^{d} \left( e^{i \theta_j} +  e^{-i \theta_j} \right) - \frac{2d}{u} \right\} d \Theta^{(d)}_{unif} \right]
\\
& = m \left( \sum_{j=1}^{d} (X_j + X_j^{-1})  - \frac{2d}{u} \right).
\end{align*}
Thus the proof is complete.
\hfill$\square$
\par
\
\par
In particular, when $d=1$ and $d=2$, we have
\begin{prop} 
\begin{align*}
{\cal L} \left( A_{RW}, T_{\infty}^1, u \right)
&= \log \left( \frac{1+\sqrt{1-u^2}}{2} \right)
= - \sum_{n=1}^{\infty} B_{2n} \frac{u^{2n}}{2n},
\\
{\cal L} \left( A_{RW}, T_{\infty}^2, u \right)
&= 
- \frac{u^2}{8} \ {}_4 F_3 \left( \frac{3}{2}, \frac{3}{2}, 1, 1 ; 2, 2, 2 ; u^2 \right) = - \sum_{n=1}^{\infty} \left( B_{2n} \right)^2 \frac{u^{2n}}{2n},
\end{align*}
for $u \in (-1,0)$, where 
\begin{align*}
B_{2n} = {2n \choose n} \left( \frac{1}{2} \right)^{2n}.
\end{align*}
\label{mitakeumi00}
\end{prop}
\par\noindent
{\bf Proof}. If $d=1$, then Eq. \eqref{ikagame01} gives
\begin{align}
{\cal L} \left( A_{RW}, T_{\infty}^1, u \right) = \log \left( - \frac{u}{2} \right) + m \left( X + X^{-1} - \frac{2}{u} \right) \quad (-1<u<0).
\label{mitakeumi0a}
\end{align}
From Eq. \eqref{lemKeyF} in Lemma \ref{lemKeyLem}, we have
\begin{align}
m \left( X + X^{-1} - \frac{2}{u} \right) = \log \left( \frac{1+\sqrt{1-u^2}}{-u} \right) \quad (-1<u<0).
\label{mitakeumi0b}
\end{align}
Combining Eq. \eqref{mitakeumi0a} with Eq. \eqref{mitakeumi0b} implies
\begin{align*}
{\cal L} \left( A_{RW}, T_{\infty}^1, u \right) = \log \left( \frac{1+\sqrt{1-u^2}}{2} \right) \quad (-1<u<0).
\end{align*}
Moreover, 
\begin{align*}
\log \left( \frac{1+\sqrt{1-u^2}}{2} \right) = - \sum_{n=1}^{\infty} B_{2n} \frac{u^{2n}}{2n} \quad (-1<u<0).
\end{align*}
When $d=2$, it follows from Eq. \eqref{ikagame01} that
\begin{align}
{\cal L} \left( A_{RW}, T_{\infty}^2, u \right) = \log \left( - \frac{u}{4} \right) + m \left( X_1 + X_1^{-1} + X_2 + X_2^{-1} - \frac{4}{u} \right) \quad (-1<u<0).
\label{omikuron01b}
\end{align}
As for the second term in the right-hand side of Eq. \eqref{omikuron01b}, if we take $c=- 4/u$ with $u \in (-1,0)$ in Lemma \ref{LemmaRV}, then we have
\begin{align}
m \left( X_1 + X_1^{-1} + X_2 + X_2^{-1} - \frac{4}{u} \right)
=\log \left( - \frac{4}{u} \right) - \frac{u^2}{8} \ {}_4 F_3 \left( \frac{3}{2}, \frac{3}{2}, 1, 1 ; 2, 2, 2 ; u^2 \right).
\label{omikuron02b}
\end{align}
Combining Eq. \eqref{omikuron01b} with Eq. \eqref{omikuron02b} gives
\begin{align*}
{\cal L} \left( A_{RW}, T_{\infty}^2, u \right) 
= - \frac{u^2}{8} \ {}_4 F_3 \left( \frac{3}{2}, \frac{3}{2}, 1, 1 ; 2, 2, 2 ; u^2 \right) \quad (-1<u<0).
\end{align*}
Furthermore we see that the definition of the generalized hypergeometric function implies 
\begin{align*}
\frac{u^2}{8} \ {}_4 F_3 \left( \frac{3}{2}, \frac{3}{2}, 1, 1 ; 2, 2, 2 ; u^2 \right) = \sum_{n=1}^{\infty} \left( B_{2n} \right)^2 \frac{u^{2n}}{2n}.
\end{align*}
Therefore the proof is complete.
\hfill$\square$
\par
\
\par
Remark that $C_{r} (A_{RW},T_{\infty}^d)$ is nothing but the return probability for RW at time $r$. In fact, if $d=1$ and $d=2$, then
\begin{align*}
C_{r} (A_{RW},T_{\infty}^1)
&= \int_{0}^{2 \pi} \left( \cos \theta \right)^r \frac{d \theta}{2 \pi} 
= \left\{ 
\begin{array}{ll}
{\displaystyle {r \choose r/2} \left( \frac{1}{2} \right)^r } & \mbox{if $r$ is even, } 
\\
\\
0 & \mbox{if $r$ is odd},
\end{array}
\right.
\\
C_{r} (A_{RW},T_{\infty}^2)
&= \left( C_{r} (A_{RW},T_{\infty}^1) \right)^2.
\end{align*}
However, if $d = 3, 4, \ldots$, then such a simple form is not known (see \cite{Norris, Spitzer}). Combining the above-mentioned observation with Proposition \ref{mitakeumi00} implies 
\begin{cor}
\begin{align*}
C_{2n} (A_{RW},T_{\infty}^1) = B_{2n}, \qquad C_{2n} (A_{RW},T_{\infty}^2)= \left( B_{2n} \right)^2,
\end{align*}
where
\begin{align*}
B_{2n} = {2n \choose n} \left( \frac{1}{2} \right)^{2n}.
\end{align*}
\end{cor}

From Proposition \ref{prop002D}, we see that
\begin{align}
{\cal L} \left( A_{RW}, T_{\infty}^d, u \right)
= \log u +  \int_{[0,2 \pi)^d} \log \Bigg( \frac{1}{u} - \frac{1}{d} e^{(d, \cos)} \left( \Theta^{(d)} \right) \Bigg) d \Theta^{(d)}_{unif}.
\label{senkyo001}
\end{align}
The second term of the right-hand side of Eq. \eqref{senkyo001} is equal to the second term of the right-hand side of Eq. (13) in \cite{GR}. If we use their notation $T_{{\cal L}}(z)$ which is {\it spanning tree generating function} (STGF) with a regular lattice ${\cal L}$, then we have
\begin{align}
T_{{\cal L}}(u) 
= \log (2d) +  \int_{[0,2 \pi)^d} \log \Bigg( \frac{1}{u} - \frac{1}{d} e^{(d, \cos)} \left( \Theta^{(d)} \right) \Bigg) d \Theta^{(d)}_{unif}.
\label{senkyo002}
\end{align}
In our setting, if we write ${\cal L}$ as $\ZM^d$, then combining Eq. \eqref{senkyo001} with Eq. \eqref{senkyo002} implies a relation between STFG on $\ZM^d$, i.e., $T_{{\ZM^d}}(u)$, and our logarithmic zeta function ${\cal L} \left( A_{RW}, T_{\infty}^d, u \right)$ as follows:
\begin{theorem}
\begin{align*}
T_{{\ZM^d}}(u) = {\cal L} \left( A_{RW}, T_{\infty}^d, u \right) - \log u + \log (2d).
\end{align*}
\label{zetaSTFG}
\end{theorem}
Moreover, following Guttmann and Rogers \cite{GR}, we define the {\it spanning tree constant} $\lambda_{\ZM^d}$ by
\begin{align*}
\lambda_{\ZM^d} = \lim_{N \to \infty} \frac{1}{N^d} \log T_{T_{N}^d}(N),
\end{align*}
if the right-hand side exists. Here $T_{T_{N}^d}(N)$ is the number of spanning trees on $T_{N}^d$. Then the following result is known (see \cite{GR}, for example) in our setting.
\begin{align}
\lambda_{\ZM^d} = \log (2d) +  \int_{[0,2 \pi)^d} \log \Bigg( 1 - \frac{1}{d} e^{(d, \cos)} \left( \Theta^{(d)} \right) \Bigg) d \Theta^{(d)}_{unif}.
\label{senkyo004}
\end{align}
Therefore combining Theorem \ref{zetaSTFG} with Eq. \eqref{senkyo004} immediately gives  
\begin{cor}
\begin{align*}
\lambda_{\ZM^d} = T_{{\ZM^d}}(1) = {\cal L} \left( A_{RW}, T_{\infty}^d, 1 \right) + \log (2d).
\end{align*}
\label{corzetaSTFG}
\end{cor}
In particular, when $d=2$, we have
\begin{cor}
\begin{align*}
\lambda_{\ZM^2} 
= {\cal L} \left( A_{RW}, T_{\infty}^2, 1 \right) + \log (4)
= \frac{4 G}{\pi},
\end{align*} 
where $G$ is the Catalan number, i.e., 
\begin{align*}
G = \sum_{n=0}^{\infty} \frac{(-1)^n}{(2n+1)^2} = 0.91596 \ldots.
\end{align*}
\label{corzetaSTFG2d}
\end{cor}

Furthermore, we obtain the following result.
\begin{theorem}
A RW on $\ZM^d$ is transient if and only if
\begin{align*}
\lim_{u \nearrow 1} u \ \frac{\partial}{\partial u} {\cal L} \left( A_{RW}, T_{\infty}^d, u \right) < \infty.
\end{align*}
\end{theorem}
\par\noindent
{\bf Proof}. For simplicity, we write the left-hand side of the following equation as the right-hand side of this:
\begin{align*}
\int_{[0,2 \pi)^d} \log \Bigg( 1 - \frac{u}{d} e^{(d, \cos)} \left( \Theta^{(d)} \right) \Bigg) d \Theta^{(d)}_{unif} = \int \log \left( 1 - u \phi (\theta) \right) d \theta.
\end{align*}
So a standard argument of the Markov process (see Spitzer \cite{Spitzer}, for example), we see
\begin{align*}
{\cal L} \left( A_{RW}, T_{\infty}^d, u \right) 
&= \int \log \left( 1 - u \phi (\theta) \right) d \theta
= - \int \sum_{n=1}^{\infty} \frac{(u \phi (\theta))^n}{n} d \theta
\\
&= - \sum_{n=1}^{\infty} \left( \int \phi^n (\theta) d \theta \right) \frac{u^n}{n}
= - \sum_{n=1}^{\infty} P_n (0,0)  \frac{u^n}{n}.
\end{align*}
Here $P_n (0,0)$ is the return probability of our RW at the origin and at $n$ step on $\ZM^d$. Thus we get
\begin{align*}
\frac{\partial}{\partial u} {\cal L} \left( A_{RW}, T_{\infty}^d, u \right)
= - \frac{1}{u} \sum_{n=1}^{\infty} P_n (0,0) u^n
= - \frac{1}{u} \sum_{n=0}^{\infty} P_n (0,0) u^n + \frac{1}{u},
\end{align*}
since $P_0 (0,0) = 1$. So we have
\begin{align*}
\sum_{n=0}^{\infty} P_n (0,0) u^n = - u \ \frac{\partial}{\partial u} {\cal L} \left( A_{RW}, T_{\infty}^d, u \right) + 1.
\end{align*}
Noting that ``RW on $\ZM^d$ is transient" if and only if ``$\lim_{u \nearrow 1} \sum_{n=0}^{\infty} P_n (0,0) u^n < \infty$" (see \cite{Spitzer}), we obtain the desired conclusion.
\hfill$\square$

\section{Conclusion \label{sec09}} 
In this paper, we presented a new relation between the Mahler measure and our zeta function for one-dimensional QWs including the Hadamard walk. Moreover we dealt with higher-dimensional QWs, that is, Grover walks. In order to clarify the comparison with QWs, we also treated the case of higher-dimensional RWs. Our results bridged between the Mahler measure and the zeta function research fields via QWs. To extend our class investigated here to a more general class would be one of the interesting problems.

\section*{Acknowledgments}
The authors are grateful to Makoto Katori and Tomoyuki Shirai for useful discussions on this subject.


\end{document}